\documentstyle[preprint,eqsecnum,aps,epsf]{revtex}
\begin{document}
\newcommand{\be}{\begin{equation}}
\newcommand{\ee}{\end{equation}}
\newcommand{\zu}{{\bf z}_1}
\newcommand{\zd}{{\bf z}_2}
\newcommand{\ztu}{\dot {\bf z}_{1 {\rm T}}}
\newcommand{\ztd}{\dot {\bf z}_{2 {\rm T}}}

\draft
\preprint{IFUM 485/FT, December 1994}
\title{Confining Bethe--Salpeter  equation from scalar QCD}

\author{N. Brambilla  and G. M. Prosperi}

\address{
Dipartimento di Fisica dell'Universit\`{a} -- Milano\\
INFN, Sezione di Milano -- Via Celoria 16, 20133 Milano
}

\maketitle
\begin{abstract}
We give a nonperturbative derivation of the Bethe--Salpeter equation
 based on the Feynman--Schwinger path integral representation of
 the one--particle propagator  in an external field. We apply the
method to the quark--antiquark system in scalar QCD and obtain
 a confining BS equation assuming  the Wilson area law in the
 straight line approximation. The result is strictly related to
 the relativistic flux tube model and to the $q \bar{q}$
semirelativistic potential.
\end{abstract}


\section{Introduction}

   Various attempts have been  done
 in the literature to apply the Bethe-Salpeter
equation to a study of the spectrum and the properties of the
mesons. The hope
was to obtain an unified and consistent description of the quark-antiquark
bound states envolving light quarks as well as heavy ones. In all such
attempts,
at our knowledge, the choice of
the kernel of the equation was purely conjectural
and only made in such a way that the successful heavy quarks potential could be
recovered in the non relativistic limit. While a kind
of derivation of the $ q \bar q $ potential from QCD can be given in the Wilson
loop context, even if at the price of not completely proved assumptions
\cite{LSG}--\cite{baker}, no similar
derivation
seems to exist for the BS equation.

  In this paper we want to consider this problem for the simplified model of
the scalar QCD.

   As it is
well known a satisfactory semirelativistic potential for the heavy
quark--antiquark system can be obtained from the assumption
  \be
i \ln W = i (\ln W)_{\rm pert} + \sigma S_{\rm min}   \,
\label{eq:iniz}
   \ee
where: $ W $ stands for the Wilson loop integral
  \be
W = {1\over 3}
 \langle {\rm Tr} {\rm P} \exp i g \{ \oint_{\Gamma} dx^{\mu} A_{\mu} \}
        \rangle  \ ;
\label{eq:loop}
  \ee
$\Gamma$ denotes a closed loop made by a quark world line ($ \Gamma _1 $), an
antiquark world line ($ \Gamma _2 $) followed in the
reverse direction and two
straight lines connecting the initial and the final points of the two world
lines; $ ({\rm ln} W)_{\rm pert}
 $ and $ S_{\rm min} $ are the perturbative evaluation
of $ \ln W $ and the minimum area enclosed by $\Gamma$ respectively;
finally the expectation value in (1.2) stands
for the functional  integration
on the gauge field alone.
 More sofisticate evaluations of $ i \ln W $ have also been
attempted  \cite{baker},\cite{sim}, \cite{ind},
 but they shall not be considered here.

   In the derivation of the potential, $ S_{\rm min} $ is further approximated
by the surface spanned by the straight lines connecting equal time points
on the quark and the antiquark worldlines, i.e. by the surface of equation
   \be
      x^0 = t   \ ,\ \ \ \ \   {\bf x} = s {\bf z}_1 (t) + (1-s) {\bf z}_2 (t)
                \ ,
\label{eq:apret}
\ee
$t$ being the ordinary time, $ {\bf z}_1 (t) $ and ${\bf z}_2 (t)$ the quark
and the antiquark positions at the time $t$, and $s$ a parameter with $ 0 \le
s \le 1 $ . The result is
   \begin{equation}
       S_{\rm min} \cong   \int_{t_{\rm i}}^{t_{\rm f}}
 dt\,  r \int_0^1 ds [1-(s \ztu
           + (1-s) \ztd)^2 ]^{1 \over 2} =
\label{eq:minform1}
        \end{equation}
where $ \dot{z}^h_{j {\rm T}} $ stands for $ ( \delta^{hk}
- {r^h r^k \over r^2}) {d{z}^k
\over dt} $ and $ {\bf r} (t) =  \zu (t) -
\zd (t) $.

   In fact it can be shown that (1.4) is correct up
to the second order in the velocities \cite{BCP94}, \cite{BP95}
 and so is  perfectly appropriate for a derivation of the
semirelativistic potential.
  For a
full relativistic extention, Eq.(1.4) cannot be correct
  as it stands,
due to the privileged role
played by the time in it. However one can try to assume (1.4)
in the
center of mass frame and then obtain implicitly
 the corresponding equation in a general
frame simply by Lorentz transformation. In this way the assumption would
become equivalent to the so called relativistic flux tube model
\cite{flux}, \cite{BP95}.

  In this paper we want to show that in scalar QCD (i.e. neglecting the spin
of the quarks) a Bethe
Salpeter (BS) equation which include confinement can be derived from (1.1) and
(1.4) in the center of mass frame. In the momentum
representation and after the factorization of the four--momentum
conservation
  $\delta$,
 the resulting kernel turns out in the form
    \be
      \hat{I}(p_1 ,  p_2; \, p_1^\prime, p_2^\prime) =
        \hat{I}_{\rm pert}(p_1 ,  p_2; \, p_1^\prime,
p_2^\prime) +
        \hat{I}_{\rm conf}(p_1,  p_2; \, p_1^\prime,
p_2^\prime)
                \
\label{eq:kern1}
          \ee
($p_1^{\prime} +p_2^{\prime} = p_1 +p_2$),
where $ \hat{I}_{\rm pert} $
 is the usual perturbative kernel which can be written,
 at the lowest order in
 the strong coupling constant,
\be
\hat{I}_{\rm pert}(p_1, p_2; p_1^\prime,p_2^\prime)= {16\over 3}  g^2
 ({p_1+p_1^\prime\over 2})^{\mu} D_{\mu \nu}(p_1^\prime-p_1)
({p_2+p_2^\prime\over 2})^{\nu} \  ,
\label{eq:pertus}
\ee
 while in the center of mass system
 $ \hat{I}_{\rm conf} $
is given by
\begin{equation}
\hat{I}_{\rm conf} (p_1 , p_2 ; p_1^\prime ,p_2^\prime)  =
  \int d^3 {\bf r} e^{i ({\bf k}^\prime- {\bf k})\cdot
 {\bf r}}  J({\bf r}, { { p}_1+p_1^\prime \over 2},
 { p_2 + p_2^\prime \over 2})
\label{eq:bsform}
\end{equation}
($ {\bf p}_1= - {\bf p}_2= {\bf k}$,
 $ {\bf p}_{1}^\prime= - {\bf p}_{2}^\prime={\bf k}^\prime $) with
 $J$ expressed as an expansion in $ {\sigma \over m^2}$
\begin{eqnarray}
J({\bf r}, q_1, q_2)& = & 4 {\sigma r \over 2}
 {1\over q_{10} +q_{20}} [ q_{20}^2 \sqrt{ q_{10}^2 -{\bf q}_{\rm T}^2}
+ q_{10}^2  \sqrt{q_{20}^2 -{\bf q}^2_{\rm T}} +\nonumber \\
& +& { q_{10}^2 q_{20}^2 \over \vert {\bf q}_{\rm T}\vert }
 ( {\rm arcsin }{ \vert {\bf q}_{\rm T} \vert \over
 \vert q_{10}\vert } + {\rm arcsin } { \vert {\bf q}_{\rm T}\vert
 \over \vert q_{20}\vert } )] + \dots
\label{eq:kerndef}
\end{eqnarray}
    Notice that by a usual
   instantaneous approximation,
from the kernel defined by (\ref{eq:kern1})-(\ref{eq:kerndef})
 one can obtain  the following
hamiltonian

\begin{eqnarray}
{\cal{H}}({\bf r}, {\bf q})& =
 &\sqrt{m_1^2+q^2} +\sqrt{m_2^2+q^2}+\nonumber \\
& +&{\sigma r\over 2}
{1\over \sqrt{m_1^2+q^2} +\sqrt{m_2^2+q^2}}
\Big \{ \sqrt{m_2^2+q^2\over m_1^2+q^2} \sqrt{m_1^2+q_{\rm r}^2}+
\sqrt{m_1^2+q^2\over m_2^2+q^2} \sqrt{m_2^2 + q_{\rm r}^2}
+\nonumber \\
&+&\Big ( {\sqrt{m_1^2+q^2} \sqrt{m_2^2+q^2} \over q_{\rm T}}
\Big )  \Big ( {\rm arcsin} {q_{\rm T}\over \sqrt{m_1^2+q^2}}
+{\rm arcsin}{q_{\rm T}\over \sqrt{ m_2^2+q^2}} \Big ) \Big \}+\dots
\nonumber \\
& +& {\cal V}_{\rm pert}({\bf r}, {\bf q})
\label{eq:gamma}
\end{eqnarray}
with an appropriate ordering prescription.
In Eq.(\ref{eq:gamma}) ${\bf q}$  stands
now for the momentum in the center of mass  frame,
  ${\bf q}_{\rm r}=
(\hat{\bf r}\cdot {\bf q}) {\hat{\bf r}}$,  $q^h_{\rm T}= (\delta^{hk}
 -\hat{r}^h \hat{r}^k ) q^k$,  while
  $ {\cal V}_{\rm pert}$ is the ordinary
 perturbative Salpeter potential. We stress
  that Eq.(\ref{eq:gamma})
is identical to the hamiltonian  for the already mentioned
relativistic flux tube model and consequently is strictly connected
 to the semirelativistic potential  as given in \cite{BCP94},
\cite{BP95} when the spin dependent part is neglected.
Notice however that the ordering in (1.9)
 corresponding to (1.7) is not
 identical to the Weyl prescription.\par
The present paper has mainly a pedagogical purpose and we have not
 done any attempt to explicitly apply the kernel (1.5)--(1.8)
to an evaluation of the spectrum and of the properties of the mesons.
 Notice however that  very interesting results have been obtained
 in this direction  by the relativistic flux tube model
\cite{olssvar} and in a sense we may consider our paper also as providing
 a more fundamental justification to that model.

   As in potential theory the starting object in our derivation is the
gauge invariant quark-antiquark propagator
   \begin{eqnarray}
& & G_4^{\rm gi}
 (x_1, x_2; \, y_1, y_2) = \langle 0 \mid \phi_2^* (x_2) U(x_2, x_1)
\phi_1 (x_1) \phi_1^* (y_1) U(y_1, y_2) \phi_2 (y_2) \mid 0 \rangle =
\nonumber\\
 & &= -{1 \over 3} {\rm Tr} \langle
U(x_2, x_1) \Delta_F^{(1)} (x_1, y_1,A) U(y_1, y_2)
 \Delta_F^{(2)} (y_2, x_2, A) \rangle  \ ,
\label{eq:proporig}
\end{eqnarray}
where $ U(b,a) = {\rm P}_{ba} \exp (ig \int_a^b dx^\mu A_\mu (x)) $ is the path
ordered gauge string (the integration path is
over the straight line joining $a$ to
$b$), while $ \Delta_F^{(1)} $ and $ \Delta_F^{(2)}$
 denote the Feynman propagators for the two
quarks in the external field $ A_\mu $. In contrast with the potential
case,
however, no semirelativistic expansion is used, but the propagators are treated
exactly using the covariant Feynman-Schwinger  path integral representation.
\par
Notice that in establishing the BS equation we have to neglect in $(i
\ln W)_{\rm pert}$ the contributions from the two extreme lines
  ${x_1 x_2} $ and $ {y_1 y_2}$ and in $S_{\rm min}$ the border
 contribution corresponding to $x_{10} \neq x_{20}$ and $y_{10}
\neq y_{20}$,
 as in the potential case.
 This is correct  for $x_1^0 -y_1^0$ and $x_2^0-y_2^0$ large
  with respect to $\vert {\bf x}_1- {\bf x}_2\vert $, $\vert {\bf y}_1
-{\bf y}_2\vert $, $ x_1^0-x_2^0$ and $y_1^0-y_2^0$. So, strictly, we
 obtain a BS equation  but for a quantity
 $G_4$ which coincides with $G_4^{\rm gi}$  in the above limit.
 This is immaterial for what concerns bound states or
 asymptotic states.  Naturally  out of the limit situation,
 $G_4$ is no longer gauge invariant as it should be
 talking of a  BS equation.
 \par
The significance of the Feynman--Schwinger
 representation in the framework of QCD
has already been appreciated in \cite{peskin} and particularly in
 \cite{sim}. With respect to Ref.\cite{sim} we have  however a
different attitude on the role of the BS equation.\par
   The plan of the paper is the following one: in Sec. 2 we illustrate our
method on the example of a one dimensional particle in a velocity
dependent potential. In Sec. 3 we apply the method to the derivation of the
Bethe--Salpeter equation for a system of two scalar particles interacting
via a scalar field  and discuss the various complications related to
the perturbative kernel.
In Sec.4 we derive the BS equation for a
quark-antiquark system in scalar QCD under the assumption discussed above and
obtain the kernel reported. Finally in Sec.5 we derive the Salpeter
potential from the BS equation.

\section{ One dimension potential Theory}

Let us consider  the model made  by a nonrelativistic particle in
 one dimension with the Hamiltonian
\be
H= {p^2\over 2m} + U(x,p)= H_0 + U
\label{eq:hamnon}
\ee
and the corresponding
 Schr\"odinger propagators
\begin{equation}
K(x,y,t) = \langle x \vert e^{-i H t} \vert y\rangle\>,
\quad \quad \quad
K_0(x,y,t)  = \langle x \vert e^{-i H_0 t} \vert y\rangle.
\label{eq:shod}
\end{equation}
{}From the operatorial identity
\be
 e^{ -i H t} = e^{- i H_0 t}
- i \int_0^t d t' e^{i H_0 (t-t')} U e^{-i H t'}\> ,
\label{eq:idop}
\ee
we obtain the equation
\be
K(x,y,t) =K_0(x,y,t) -i \int_0^t d t' \int d \xi \int d \eta
 K_0(x,\xi, t-t') \langle \xi \vert U\vert \eta\rangle
 K(\eta,y,t').
\label{eq:intpropnon}
\ee
 which is somewhat  analogous
to the nonhomogeneous
Bethe-Salpeter equation
 in the configuration space.\par
We want to derive Eq.(\ref{eq:intpropnon}) by the
path--integral
formalism.\par
 Let us be for definiteness
\be
U= V(x) + (W(x) p^2)_{\rm ord}\  ,
\label{eq:ordnon}
\ee
where $(\>)_{\rm ord}$ stands for some ordering prescription.
In terms of path--integral we can write in the phase space
\be
K(x,y,t) = \int_y^x {\cal D}z {\cal D}p \exp \{i \int_0^t dt'
 [p' \dot{z'} - {p^{'2}\over 2 m} - V(z')- W(z')p^{'2}] \}
\label{eq:pathmomnon}
\ee
with $z^\prime= z(t^\prime)$, $p^\prime= p(t^\prime)$,
$\dot{z}^\prime={ d z(t^\prime) \over d t^\prime}$.
In Eq.(\ref{eq:pathmomnon})
the functional ``measures'' are supposed to be defined by
\begin{eqnarray}
 {\cal D} z  & = & ({m\over 2 \pi i \varepsilon })^{  N\over 2}
  d z_1 \dots  d z_{N-1}, \quad \qquad \qquad
 {\cal D} p   =  ({i \varepsilon\over 2 \pi m})^{  N\over 2}
  d p_1 \dots  d p_{N-1} d p_N \nonumber \\
& & {\cal D} z  {\cal D} p  =  ({1\over 2 \pi })^{  N}
 d p_1 d z_1 \dots d p_{N-1} d z_{N-1}
 d p_N
\label{eq:monodim}
\end{eqnarray}
where $\varepsilon={t\over N}$,
the limit $N\to \infty$ is understood and the end points $x$
 and $y$ stand for the condition $z_0=y$, $z_N=x$.
 As  well known (see e.g. \cite{lan})
 the ordering prescription is  concealed under
 the particular discretization
 adopted in the limit procedure implied in the definition
of (\ref{eq:pathmomnon}).\par
Let us first  consider  the case $W=0$. In this case
 there is no ordering problem and
 it is possible to calculate explicitly the
 integral in $p$ in (\ref{eq:pathmomnon})
obtaining the path-integral representation
  in the configuration space
\begin{equation}
 K(x,y,t)  = \int_y^x {\cal D}z e^{i \int_0^t dt' (m {\dot{z}^{'2}\over 2 }
- V(z'))}.
\label{eq:duenove}
\end{equation}
\noindent
Then using the identity
\be
 e^{-i \int_0^t d t' V(z')} = 1-i \int_0^t d t' V(z')
 e^{-i\int_0^{t'} dt''
V(z'')}
\label{eq:idnonrel}
\ee
 one obtains
\be
K(x,y,t)= K_0(x,y,t) -i \int_0^t dt' \int_y^x {\cal D}z
 V(z') e^{ i \int_{t'}^t d t'' m { \dot{z}^{''2}\over 2 }+
i \int_0^{t'} d t''  m{ \dot{z}^{''2} \over 2 }
  -i
 \int_0^{t'} dt'' V(z'')}
\label{eq:dued}
\ee
which,  taking into account that
\be
\int_y^x {\cal D} z\dots  =\int d \xi \int_{\xi}^x {\cal D}z \int_y^{\xi}
{\cal D} z \dots
\label{eq:misspez}
\ee
(having identified $\xi=z(t')$),  can be rewritten
 in the form
(\ref{eq:intpropnon}) with
 $\langle\xi \vert U\vert \eta\rangle=
V(\xi)
\delta(\xi-\eta)$.\par
In the general case $W(x) \neq 0$  it is convenient
  to work  with the  original
  path-integral representation
 in the phase space (\ref{eq:pathmomnon}) and it is necessary
 to use discretized
 expressions explicitly.
 For  Weyl ordering in Eq.(\ref{eq:ordnon}) the correct discretization
 is the   mid--point  one and
 we can write
\begin{eqnarray}
& &K(x,y,t)= {1\over (2 \pi)^N} \int d p_N d z_{N-1} d p_{N-1}
\dots d z_1 d p_1 \nonumber \\
& & \exp i \sum_{n=1}^{N} \{ p_n (z_n-z_{n-1})
-\varepsilon [ {p_n^2 \over 2 m}+ V({x_n +x_{n-1}\over 2}) +
 W({x_n+x_{n-1}\over 2}) p_n^2 ] \}.
\label{eq:discrprop}
\end{eqnarray}
 Then it is convenient  to introduce the Fourier Transform
 of $K(x,y,t)$
\be
\tilde{K}(k,q,t)= \int dx \int dy  e^{-i kx}
 {K}(x,y,t) e^{iqy}
\label{eq:four}
\ee
and to use the discrete counterpart of (\ref{eq:idnonrel})
\begin{eqnarray}
& &\exp{ (-i \varepsilon) \sum_{n=1}^N [V({x_n+x_{n-1}\over 2})
+W({x_n+x_{n-1}\over 2}) p_n^2 ]}=
1- i\varepsilon \sum_{R=1}^{N} \{ [ V({x_R+x_{R-1}\over 2}) +
 W({x_R+x_{R-1} \over 2}) p_R^2 ]\nonumber \\
& & \cdot
\exp{(-i\varepsilon)} \sum_{r=1}^{R-1} [ V({x_r +x_{r-1}
\over 2}) + W({x_r+x_{r-1} \over 2}) p_r^2 ].
\label{eq:discrid}
\end{eqnarray}
Replacing (\ref{eq:discrid}) in (\ref{eq:discrprop}) and
 Fourier transforming we have
\begin{eqnarray}
& & \tilde{K}(k,q,t)= \tilde{K}_0(k,q,t) -i\varepsilon {1\over (2\pi)^N}
\sum_{R=1}^{N} \int dz_N dp_N \dots dp_1 dz_0\,
 e^{-i kz_N} \exp i \sum_{n=R}^N p_n (z_n-z_{n-1})\nonumber \\
& &[V({z_R+z_{R-1}\over 2}) +W({z_R +z_{R-1}\over 2}) p_R^2 ]
 \exp i \sum_{n=1}^{R-1} [p_n(z_n-z_{n-1}) -\nonumber \\
& &\varepsilon ({p_n^2 \over 2m} +V({z_n+z_{n-1}\over 2})
+W({z_n +z_{n-1}\over 2}) p_n^2)] e^{i qz_0}=\nonumber \\
& & = \tilde{K}_0(k,q,t) -i \varepsilon \sum_{R=1}^{N} {1\over (2 \pi)^2}
\Big \{\int d p_{R+1} \int d p_{R-1} {1\over (2 \pi)^{N-R-1}}
 \int dz_N dp_N \dots dp_{R+2} dz_{R+1}  e^{-ik z_N}\nonumber\\
& & \exp \{i \sum_{n=R+2}^N p_n (z_n-z_{n-1}) \} e^{i p_{R+1}
z_{R+1}}\Big \}
\nonumber \\
&&
\Big \{{1\over 2 \pi} \int d z_R dp_R dz_{R-1} e^{-i (p_{R+1}-p_R)z_{R}}
 [V({z_R+z_{R-1} \over 2}) + W({z_R +z_{R-1}\over 2}) p_R^2 ]
 e^{i (p_{R-1}-p_R) z_{R-1}} \Big \}\nonumber \\
&& \Big \{{1\over (2 \pi)^{R-2}} \int dz_{R-2} \dots dz_0 e^{-i p_{R-1}
 z_{R-2}} \exp i \sum_{n=1}^{R-2} \{ p_n (z_n-z_{n-1})\nonumber \\
& &-\varepsilon [{p_n^2 \over 2m } + V({z_n+z_{n-1} \over 2})
+W({z_n+ z_{n-1}\over 2}) p_n^2 ]\}   e^{iq z_0}\Big \}
\label{eq:monster}
\end{eqnarray}
 which in the continuous limit  reads
\be
\tilde{K}(k,q,t)=\tilde{K}_0(k,q,t) -i \int_0^t dt^\prime \int {dk^\prime
\over 2 \pi} \int {d q^\prime\over
2\pi} \tilde{K}_0(k, k^\prime,t- t^\prime)
 \tilde{I}_{\rm W}(k^\prime,q^\prime)\tilde{K}(q^\prime, q, t^\prime)
\label{eq:cont}
\ee
with

\begin{eqnarray}
& &\tilde{I}_{W}(k,q)= {1\over 2 \pi} \int dz^{\prime\prime}
 dp dz^\prime e^{-ikz^{\prime\prime}}
 e^{ipz^{\prime\prime}} [V({z^{\prime\prime}
+z^{\prime}\over 2}) + W({z^{\prime\prime}+z^\prime \over 2}) p^2 ]
e^{-ipz^\prime}
e^{iqz^\prime}=\nonumber\\
& & =\tilde{V}(k-q) +\tilde{W}(k-q) ({k+q\over 2})^2\equiv
\langle k \vert (V(x) +{1\over 4} \{ p, \{ p,W(x)\}\}) \vert q\rangle.
\label{eq:cont2}
\end{eqnarray}
\noindent
Eq.(\ref{eq:cont}) is equivalent  to Eq.(2.4) and the kernel reduces to
the Fourier transform of the case $W(x)=0$.\par
Had we considered  the   symmetric
ordering in Eq.(\ref{eq:ordnon}) we should have replaced
in (\ref{eq:discrid}) and (\ref{eq:monster})
 $\big [V({x_n +x_{n-1}\over 2}) + W({x_n + x_{n-1}\over 2}) p_n^2
\big ] $
 with $\big [ {1\over 2} ( V(x_n) +V(x_{n-1})) +{1\over 2} (W(x_n)+W(x_{n-1}))
p_n^2\big ]$ and the result would be
\begin{eqnarray}
& &\tilde{I}_{S}(k,q)= {1\over 2 \pi} \int dz^{\prime\prime}
 dp\, dz^\prime e^{-ikz^{\prime\prime}}
 e^{ipz^{\prime\prime}} [{V(z^{\prime\prime}) +V(z^\prime)\over
 2} + {W(z^{\prime\prime}) +W(z^\prime) \over 2} p^2 ]
e^{-ipz^\prime}
e^{iqz^\prime}=\nonumber\\
& &= \tilde{V}(k-q) +\tilde{W}(k-q) {k^2+q^2 \over 2}\equiv
\langle k \vert (V(x) +{1\over 2} \{ p^2,W(x)\}) \vert q\rangle
\label{eq:cont3}
\end{eqnarray}

\section{Spinless particles interacting
throught a scalar field}
Let us consider  two scalar ``material'' fields $\phi_1$ and $\phi_2$
interacting  through a third scalar field $A$ with the coupling
 ${1\over 2} (g_1 \phi_1^2  A + g_2 \phi_2^2 A)$.
Then,  after integration over $\phi_1$ and $\phi_2$,
the full one particle propagator can be written as
\begin{equation}
G_{2}^{(j)}(x-y)=  \langle 0 \vert {\rm T}\phi_j (x) \phi_j (y)\vert
0 \rangle= \langle i\Delta_{\rm F}^{(j)}(x, y; A)\rangle \equiv
 {\int {\cal D}A e^{i S_0(A)} M(A) i\Delta_{\rm F}^{(j)}(x,y; A) \over
 \int {\cal D}A e^{i S_0(A)} M(A)}
\label{eq:propscal}
\end{equation}
where  $\Delta_{\rm F}^{(j)}(x,y,A)$ is the propagator for the
particle $j$
 in the external
 field $A$,
 $S_0(A)$ is the free
 action for the field $A$ and the determinantal factor $M(A)$
 comes from the integration of the fields $\phi_j$
\begin{eqnarray}
& & M(A)= \prod_{j=1,2}
\Big [{ {\rm det} (\partial^{\mu} \partial_{\mu} +m_j^2 - g_j A)
\over {\rm det} (\partial^{\mu} \partial_{\mu} +m_j^2)}\Big ] ^{-{1\over
2}}=\nonumber \\
& & =1- {1\over 2} \sum_{j=1,2} \big \{- g_j \int d^4 x A(x)
\Delta_{\rm F}^{(j)}(0) -{1\over 2} g_j^2 \int d^4 x d^4 y A(x)
 \Delta_{\rm F}^{(j)}(x-y) A(y) \Delta_{\rm F}^{(j)}(y-x) + \dots \}
\label{eq:det}
\end{eqnarray}
\indent The covariant
 Feynman-Schwinger representation for $\Delta_{\rm F}^{(j)}$ reads
\begin{eqnarray}
\Delta_{\rm F}^{(j)}(x,y; A)& =&
-{ i\over 2} \int_0^{\infty} d \tau
 \int_{y}^{x} {\cal D} z {\cal D} p\,
{\rm exp}\, \big \{  i\int_0^{\tau} {\rm d} \tau^\prime [ - p_{\mu \prime}
 \dot{z}^{\prime\mu} +{1\over 2} p^{\prime}_{\mu} p^{\prime \mu }
-{1\over 2} m_j^2 +{1\over 2} g_j A(z') ] \big \}
\nonumber\\
& =& -{ i\over 2} \int_0^{\infty} d \tau
 \int_{y}^{x} {\cal D} z\,
{\rm exp}\,\big \{ - i\int_0^{\tau} {\rm d} \tau^\prime  {1\over 2} [
 ( \dot{z}^{\prime 2} +m_j^2) - g_j  A (z^\prime) ] \big \},
\label{eq:propcov}
\end{eqnarray}
 where the path integrals are understood to be extended over all
 paths
 $z^{\mu}= z^{\mu}(\tau^\prime)$ connecting $y$ with $x$ expressed in
terms of an arbitrary  parameter $ \tau^\prime$
with $ 0\le \tau'
\le \tau$. In Eq.(\ref{eq:propcov}) $z^\prime$ stands for $z(\tau^\prime)$,
$p^\prime$ for $ p(\tau^\prime)$, $\dot{z}^\prime$
for ${dz(\tau^\prime)
\over d \tau^\prime}$ and the ``functional measures'' are assumed to be
defined as
\begin{eqnarray}
 {\cal D} z  & = & ({1\over 2 \pi i \varepsilon })^{ 2 N}
  d^4 z_1 \dots  d^4 z_{N-1}, \quad \qquad \qquad
 {\cal D} p   =  ({i \varepsilon\over 2 \pi })^{ 2 N}
  d^4 p_1 \dots  d^4 p_{N-1} d^4 p_N \nonumber \\
& & {\cal D} z  {\cal D} p  =  ({1\over 2 \pi })^{ 4 N}
 d^4 p_1 d^4 z_1 \dots d^4 p_{N-1} d^4 z_{N-1}
 d^4 p_N.
\label{eq:misfunz}
\end{eqnarray}
Replacing  Eq.(\ref{eq:propcov}) in (\ref{eq:propscal})
 we obtain
\begin{equation}
G_2^{(j)} (x-y)= {1\over 2} \int_0^{\infty}d \tau \int_y^x {\cal D} z
 \exp \{ -{i\over 2} \int_0^\tau d \tau' (\dot{z}'^2 +m_j^2)\}
\langle \exp {i g_j \over 2} \int_0^{\tau} d \tau' A(z')\rangle
\label{eq:propppscal}
\end{equation}
where, suppressing the tadpole term in (3.2),
\begin{eqnarray}
 \langle \exp {i g_j \over 2} \int_0^{\tau} d \tau^\prime
A(z')\rangle & &  = \exp {i g_j^2\over 4} \int_0^{\tau} d \tau^\prime
 \int_0^{\tau^\prime} d\tau^{\prime \prime} \big [
 D_{\rm F}(z^\prime-z^{\prime \prime}) +\nonumber \\
& & + \sum_{i} {g_i^2\over 2}
\int d^4 \xi \int d^4 \eta D_{\rm F}(z^{\prime} -\xi) (\Delta_{\rm
F}^{(i)}(\xi-\eta))^2 D_{\rm F}(\eta- z^{\prime \prime}) +\dots \big ]
\label{eq:av1}
\end{eqnarray}
\indent Now let us consider   the two particle propagator
 \begin{eqnarray}
& & G_4(x_1,x_2; y_1, y_2) = \langle 0\vert {\rm T} \phi_1(x_1)
\phi_2(x_2) \phi_1(y_1) \phi_2(y_2) \vert 0\rangle=
\langle G_2^{(1)}(x_1, y_1;A)  G_2^{(2)} (x_2, y_2;A)
\rangle = \nonumber \\
& &
 \quad \qquad \quad
= ({1\over 2})^2 \int_0^{\infty} d \tau_1 \int_0^{\infty} d \tau_2
 \int_{y_1}^{x_1} {\cal D}z_1 \int_{y_2}^{x_2} {\cal D} z_2\times
\nonumber \\
& &\exp {-i \over 2} \{ \int_0^{\tau_1} d \tau_1' (\dot{z}_1^{'2}
+m_1^2) +\int_0^{\tau_2} d \tau_2' (\dot{z}_2^{'2}+m_2^2)\}
\langle \exp {i\over 2} \{ g_1 \int_0^{\tau_1} d \tau_1'
 A(z_1') +g_2 \int_0^{\tau_2} d \tau_2' A(z_2')\}
\rangle
\label{eq:g4}.
\end{eqnarray}
In this case
\begin{eqnarray}
& & \langle \exp {i\over 2} \{ g_1 \int_0^{\tau_1} d \tau_1'
 A(z_1') +g_2 \int_0^{\tau_2} d \tau_2' A(z_2')\}
\rangle =\nonumber \\
& & \exp \sum_{j=1,2} {i g^2_j \over 4} \int_0^{\tau_j} d \tau_j^{\prime}
 \int_0^{\tau_j^\prime} d \tau_j^{\prime\prime} \big [
 D_{\rm F}(z^{\prime}_j - z^{\prime \prime}_j)
+\sum_{i=1,2} {g_i^2\over 2}
\int d^4 \xi \int d^4 \eta D_{\rm F}(z^{\prime}_i -\xi)\times
\nonumber \\
& &  (\Delta_{\rm
F}^{(i)}(\xi-\eta))^2 D_{\rm F}(\eta- z^{\prime \prime}_i)
 +\dots \big ]
 + {i g_1 g_2 \over 4} \int_0^{\tau_1} d\tau_1^{\prime}
 \int_0^{\tau_2} d\tau_2^{\prime} \big [ D_{\rm F}(z_1^\prime
-z_2^\prime) +\nonumber \\
& & +\sum_{i=1,2} {g_i^2\over 2}
\int d^4 \xi \int d^4 \eta D_{\rm F}(z^{\prime}_i -\xi)  (\Delta_{\rm
F}^{(i)}(\xi-\eta))^2 D_{\rm F}(\eta- z^{\prime \prime}_i)
 +\dots \big ]
\label{eq:treottobis}
\end{eqnarray}
If we replace $M(A)$ by $1$ in (\ref{eq:propscal}) and
 (\ref{eq:g4}), i.e., if we retain
only the lowest order terms in the exponent in (\ref{eq:treottobis})
 (quenched
approximation), we have no ``material'' fields loops in the evaluation
 of the gauge field average (\ref{eq:av1}) and (\ref{eq:treottobis})
and then we can write exactly
\begin{eqnarray}
& & G_4(x_1,x_2; y_1, y_2)
 =  ({1\over 2})^2 \int_0^{\infty} d \tau_1 \int_0^{\infty} d \tau_2
 \int_{y_1}^{x_1} {\cal D}z_1 \int_{y_2}^{x_2} {\cal D} z_2
\nonumber \\
& & \exp {-i \over 2} [ \int_0^{\tau_1} d \tau_1' (\dot{z}_1^{'2}
+m_1^2) -   {g_1^2\over 2} \int_0^{\tau_1} d \tau_1'
 \int_0^{\tau_1'} d \tau_1'' D_{\rm F}(z_1'-z_1'')]\times
\nonumber\\
& & \exp {- i \over 2} [
\int_0^{\tau_2} d \tau_2' (\dot{z}_2^{'2}+m_2^2)\}
 -{g_2^2\over 2} \int_0^{\tau_2} d \tau_2' \int_0^{\tau_2'} d \tau_2''
 D_{\rm F}(z_2'-z_2'') ]\times \nonumber \\
& & \exp  {i g_1 g_2\over 4} \int_0^{\tau_1} d \tau_1'
 \int_0^{\tau_2} d \tau_2' D_{\rm F}(z_1' -z_2').
\label{eq:g4bis}
\end{eqnarray}
Proceding as in Sec.2, using
the identity
\begin{eqnarray}
& & \exp {ig_1 g_2 \over 4} \int_0^{\tau_1} d \tau_1'
 \int_0^{\tau_2} d \tau_2' D_{\rm F}(z_1'-z_2')=  \nonumber \\
 & &=1 + {i g_1 g_2 \over 4} \int_0^{\tau_1} d \tau_1'
 \int_0^{\tau_2} d \tau_2' D_{\rm F}(z_1'-z_2') \exp
 [ {i g_1 g_2 \over 4} \int_0^{\tau_1'}
 d \tau_1'' \int_0^{\tau_2} d \tau_2'' D_{\rm F}(z_1''-z_2'')]
\label{eq:id}
\end{eqnarray}
( analogous to (\ref{eq:idnonrel}))
and Eq.(\ref{eq:propppscal}),
  we obtain after some manipulations
\begin{eqnarray}
& & G_4(x_1, x_2; y_1, y_2) = G_2(x_1-y_1) G_2(x_2-y_2)+
   {i g_1 g_2 \over 4} \nonumber \\ & &
 ({1 \over 2})^2 \int_0^{\infty} d \tau_1'
 \int_{\tau_1'}^{\infty} d \tau_1 \int_0^{\infty} d \tau_2'
\int_{\tau_2'}^{\infty} d \tau_2
\int d^4z_1' \int d^4z_2'
\int_{z_1'}^{x_1}  {\cal D} z_1
\int_{z_2'}^{x_2} {\cal D} z_2 \int_{y_1}^{z_1'} {\cal D} z_1
 \int_{y_2}^{z_2'} {\cal D} z_2\, D_{\rm F}(z_1'-z_2') \nonumber \\
& &  \exp {i\over 2}
 \big \{ \int_{\tau_1'}^{\tau_1} d \tau_1'' [ -\dot{z}_1^{''2}-m_1^2
  +{g_1^2 \over 2} \int_{\tau_1'}^{\tau_1''} d \tau_1'''
D_{\rm F}(z_1''-z_1''')] +
 \int_{\tau_2'}^{\tau_2} d \tau_2'' [ - \dot{z}_2^{''2}-m_2^2
 +{g_2^2 \over 2} \int_{\tau_2'}^{\tau_2''} d \tau_2'''\nonumber \\
& & D_{\rm F}(z_2''-z_2''')]\big \}
 \times
\exp{i\over 2} \big \{\int_{0}^{\tau_1'} d \tau_1'' [ - \dot{z}_1^{''2}-m_1^2
 +{g_1^2 \over 2} \int_{0}^{\tau_1''} d \tau_1'''
D_{\rm F}(z_1''-z_1''')]
+\int_{0}^{\tau_2'} d \tau_2'' [  -\dot{z}_2^{''2}-m_2^2+
\nonumber \\
& &\quad  +{g_2^2 \over 2} \int_{0}^{\tau_2''} d \tau_2'''
D_{\rm F}(z_2''-z_2''')]
 +{g_1 g_2 \over 2} \int_0^{\tau_1'} d \tau_1''
 \int_0^{\tau_2'} d \tau_2'' D_{\rm F}
(z_1''-z_2'')\big \} \times \nonumber \\
& & \exp i\big \{ {g_1^2 \over 4}
 \int_{\tau_1'}^{\tau_1} d \tau_1'' \int_0^{\tau_1'} d \tau_1'''
 D_{\rm F}(z_1''-z_1''') +
{g_2^2 \over 4}
 \int_{\tau_2'}^{\tau_2} d \tau_2'' \int_0^{\tau_2'} d \tau_2'''
 D_{\rm F}(z_2''-z_2''') +\nonumber \\
& & \quad \quad \quad \quad + {g_1 g_2 \over 4}
\int_0^{\tau_1'} d \tau_1'' \int_{\tau_2'}^{\tau_2} d \tau_2''
 D_{\rm F}(z_1''-z_2'')\big \}.
\label{eq:orror}
\end{eqnarray}
Then,  if we denote by $L_1$ the last exponential in this equation and
replace it by 1,
we obtain immediately
the Bethe--Salpeter equation
\begin{eqnarray}
& & G_4(x_1, x_2, y_1, y_2)  =  G_2(x_1-y_1) G_2(x_2-y_2)+\nonumber \\
& &   -i \int d^4 z_1 \int d^4 z_2 \int d^4 z_1' \int d^4 z_2'
G_2(x_1-z_1) G_2(x_2-z_2) I(z_1,z_2; z_1',z_2') G_4(z_1',z_2',y_1,
y_2)
\label{eq:bethesconf}
\end{eqnarray}
with  the ladder approximation kernel
\begin{equation}
I(z_1,z_2,z_1^\prime,z_2^\prime)= - g_1 g_2 D_{\rm
F}(z_1^\prime-z_2^\prime)
\delta^4(z_1-z_1^\prime) \delta^4(z_2-z_2^\prime).
\label{eq:ladder}
\end{equation}
On the contrary if  we introduce in (\ref{eq:orror}) the entire
 expansion  of $L_1$  in the second line of the  equation
 beside $D_{\rm F}(z_1^\prime-z_2^\prime)$ we obtain additional terms
 of the type ${g_i^2\over 4} \int_{\tau_i^\prime}^{\tau} d
\tau_i^{\prime\prime} \int_0^{\tau_i} d\tau_i^{\prime\prime\prime}
D_{\rm F} (z_i^{\prime \prime} -z_i^{\prime\prime\prime}) D_{\rm
F}(z_1^\prime -z_2^\prime)$, $ g_1 g_2 \int_{\tau_1^\prime}^{\tau}
 d \tau_1^{\prime\prime} \int_0^{\tau_2^\prime} D_{\rm
F}(z_1^{\prime\prime}-z_2^{\prime\prime}) D_{\rm
F}(z_1^\prime-z_2^\prime)$ etc.. As a consequence (see App.A for
 details) we reobtain
Eq.(\ref{eq:bethesconf}) but with the kernel
\begin{eqnarray}
& & I(z_1,z_2,z_1',z_2') = \nonumber \\
& & \quad = -g_1 g_2  D_{\rm F}(z_1'-z_2')
\delta^4(z_1'-z_1) \delta^4(z_2'-z_2)
+i g_1^3 g_2 \int d^4\xi_1 D_{\rm F}(z_1-z_1')
 G_2(z_1-\xi_1)
 G_2(\xi_1-z_1^\prime)\nonumber \\
& &
 D_{\rm F}(\xi_1-z_2) \delta^4(z_2-z_2')+ i
g_1 g_2^3 \int d\xi_2 \delta^4(z_1-z_1')
D_{\rm F}(z_1-\xi_2)
  G_2(z_2-\xi_2)
 G_2(\xi_2-z_2')\nonumber \\
& &   D_{\rm F}(z_2-z_2')
+ i g_1^2 g_2^2 D_{\rm F}(z_1-z_2^\prime) G_2(z_1-z_1')
D_{\rm F}(z_2-z_1') G_2(z_2-z_2')
+\dots
\label{eq:kerntant}
\end{eqnarray}
or, graphically:
\epsfxsize=16truecm
\epsffile{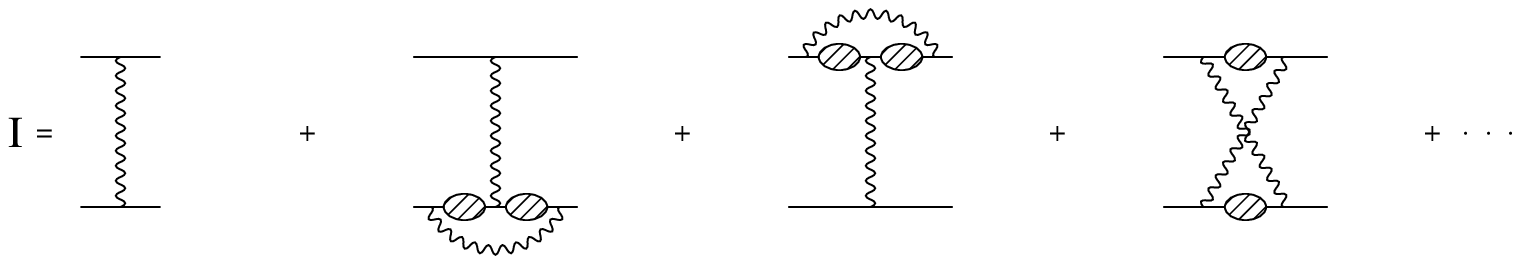}
\centerline{\it{Figure caption}.}
\vskip 1truecm
Finally to go beside  the quenched approximation and
  to take into account additional terms in Eq.(\ref{eq:propscal}),
 (\ref{eq:treottobis}) amounts to insert $\phi_1 \bar{\phi}_1$
and $\phi_2 \bar{\phi}_2$  loops
in all
possible  way inside the graph.\par
Had we used Eq.(\ref{eq:propppscal})
 in the phase space and a method analogous to that
 employed starting from Eq.(2.12) we would have
obtained the
Bethe--Salpeter equation in the momentum space
\begin{eqnarray}
& &  \tilde{G}_4 (p_1, p_2 ; p_1^\prime,p_2^\prime) = \tilde{G}^{(1)}_2
(p_1, p_1^\prime) \tilde{G}^{(2)}_2(p_2 , p_2^\prime)  \nonumber \\
 & & -i  \int {d^4 k_1^\prime \over (2\pi)^4}
{d^4 k_2^\prime \over (2 \pi)^4} {d^4 k_1\over (2 \pi)^4}
 {d^4 k_2 \over (2 \pi)^4} \tilde{G}_2 (p_1 ,k_1)
 \tilde{G}_2(p_2 ,k_2) \tilde{I}(k_1 , k_2 ;
 k_1^\prime,k_2^\prime)
\tilde{G}_4(k_1^\prime,k_2^\prime;p_1^\prime,p_2^\prime).
\label{eq:bspur}
\end{eqnarray}
with
\begin{eqnarray}
& & \tilde{I}(p_1,p_2, p_1^\prime , p_2^\prime) = \int d^4 z_1
 d^4 z_2 \int {d^4 k_1 \over (2  \pi)^4}
 {d^4 k_2\over (2 \pi)^4} \int d^4
z_1^\prime
 d^4 z_2^\prime e^{-i (p_1 -k_1) z_1} e^{-i
(p_2-k_2)z_2}\nonumber \\
& &   4  g_1 g_2 D({z_1+ z_1^\prime\over 2} -
 {z_2+z_2^\prime\over 2}) \,
 e^{i (p_1^\prime-k_1) z_1^\prime} e^{i (p_2^\prime-k_2) z_2^\prime}
\end{eqnarray}
where for definiteness we have assumed the mid--point prescription
 in the discretized  form of (3.9) even if immaterial in this case.
Then, introducing the total momentum $P=p_1+p_2$ and the relative
 momentum
 $q= {m_2\over m_1+m_2} p_1 -{m_1 \over m_1+m_2} p_2$,
 defining
\begin{eqnarray}
& & \tilde{G}_2(p ,p^\prime) = (2 \pi)^4 \delta^4(p-p^\prime) \hat{G}_2(p)
\quad \quad  \tilde{G}_4(p_1,p_2;p_1^\prime,p_2^\prime)= (2 \pi)^4
 \delta^4(P-P^\prime) \hat{G}_4(q,q^\prime;P)\nonumber \\
& & \quad \quad \quad \quad
\tilde{I}(p_1, p_2; p_1^\prime, p_2^\prime) = (2 \pi )^4
\delta^4(P-P^\prime) \hat{I}(q, q^\prime; P)
\label{eq:capdef}
\end{eqnarray}
and factorizing the total momentum  conservation delta,
 we can  write in conclusion
\be
\hat{G}_4(q;q^\prime,P) = (2 \pi)^4 \delta^4(q-q^\prime)
 \hat{G}_2^{(1)}(q)
\hat{G}_2^{(2)}(-q) -i \int {d^4 k\over (2\pi)^4}
\hat{G}_2^{(1)}(q)
\hat{G}_2^{(2)}(-q) \hat{I}(q ,k, P) \hat{G}_4(k,q^\prime,P)
\label{eq:momrid}
\ee
\noindent
and
in the ladder approximation we  would have obtained
\be
\hat{I}(q,q^\prime;P) = -g_1 g_2 D_{\rm F}(q-q^\prime)
\label{eq:ladmom}
\ee

\section{Scalar QCD}
Let us come to the scalar QCD characterized by the lagrangian
\be
L= (D_{\rho}\phi)^{*} D^{\rho} \phi - m^2 \phi^{*} \phi - {1\over 2 }
{\rm Tr} F_{\mu\nu} F^{\mu \nu} + L_{\rm GF}
\label{eq:lagqcd}
\ee
where $D_{\rho}= \partial_{\rho} +i e A_{\rho}$ and
 $L_{\rm GF}$ is the gauge fixing term. In this case an equation
 analogous to (\ref{eq:propcov}) can be obtained
\begin{eqnarray}
\Delta_{\rm F}^{(j)}(x,y; A_{\mu}) &=& -{i\over 2} {\rm P}_{xy}
\int_0^{\infty} d \tau \int_y^x {\cal D}z \exp \{ -i
\int_0^{\tau} d\tau^\prime [ {1\over 2} (\dot{z}^{\prime 2}
+m_j^2) - \dot{z}^{\mu\prime} A_{\mu}(z^\prime)]\}\nonumber \\
 \Delta_{\rm F}^{(j)}(y,x; A_{\mu}) &=& -{i\over 2} {\rm P}_{yx}
\int_0^{\infty} d \tau \int_y^x {\cal D}z \exp \{ -i
\int_0^{\tau} d\tau^\prime [ {1\over 2} (\dot{z}^{\prime 2}
+m_j^2) +\dot{z}^{\mu\prime} A_{\mu}(z^\prime)]\}\ .
\label{eq:propord}
\end{eqnarray}
 Using (\ref{eq:propord})  in (\ref{eq:proporig}) we have
\begin{eqnarray}
G_4^{q \bar{q}}(x_1,x_2; y_1,y_2) &=&
 ({1\over 2})^2 \int_0^{\infty} d \tau_1 \int_0^{\infty} d\tau_2
 \int_{y_1}^{x_1} {\cal D}z_1 \int_{y_2}^{x_2} {\cal D}z_2
\exp {-i\over 2} \{ \int_0^{\tau_1} d \tau_1' (m_1^2 +\dot{z}_1^{'2})
+\nonumber \\
&& + \int_0^{\tau_2} d \tau_2' (m_2^2 +\dot{z}_2^{'2}) \}
\,  {1\over 3}
\langle {\rm Tr} {\rm P} \exp [ ig \oint_{\Gamma} d z^{\mu} A_{\mu}(z)]
\rangle=\nonumber \\
& & = ({i\over 2})^2 \int_0^{\infty} d \tau_1 \int_0^{\infty} d\tau_2
 \int_{y_1}^{x_1} {\cal D}z_1 \int_{y_2}^{x_2} {\cal D}z_2
\exp {i S^{q \bar{q}}}
\label{eq:propwilson}
\end{eqnarray}
$S^{q \bar{q}}$  being a kind of effective $q \bar{q}$ action.
Then, in the quenched approximation, keeping $(\ln W)_{\rm pert}$
 at the lowest order and rewriting appropriately
  Eq.(\ref{eq:minform1}), one has
\begin{eqnarray}
& &i \ln W= i \ln \langle {1\over 3} {\rm Tr} {\rm P} \exp ig \oint_{\Gamma}
dz^{\mu}
 A_{\mu}(z)\rangle ={4\over 3} g^2 \int_0^{\tau_1} d \tau_1^\prime
\int_0^{\tau_2} d\tau_2^\prime D_{\mu \nu}(z_1^\prime-z_2^\prime)
 \dot{z}_1^{\prime \mu}
\dot{z}_2^{\prime \nu}+ \nonumber \\
& & +{2\over 3} g^2 \int_0^{\tau_1} d \tau_1^\prime
\int_0^{\tau_1} d\tau_1^{\prime\prime}
 D_{\mu \nu}(z_1^\prime-z_1^{\prime\prime}) \dot{z}_1^{\prime\mu}
\dot{z}_1^{\prime\prime\nu}+
{2\over 3} g^2 \int_0^{\tau_1} d \tau_2^\prime
\int_0^{\tau_2} d\tau_2^{\prime\prime} D_{\mu
\nu}(z_1^\prime-z_1^{\prime\prime}) \dot{z}_2^{\prime\mu}
\dot{z}_2^{\prime \prime\nu}\nonumber \\
& &  +\sigma
  \int_0^{\tau_1} d \tau_1^\prime \int_0^{\tau_2}
 d \tau_2^\prime\,  \delta( z_{10}^{\prime} -z_{20}^{\prime})
 \vert {\bf z}_1^{\prime} -{\bf z}_2^{\prime}\vert
\int_0^1 d s \big \{ {\dot{z}_{10}^{\prime 2}} {\dot{z}_{20}^{\prime 2}} -
 (s \dot{\bf z}_{1{\rm T}}^{\prime} \dot{z}_{20}^{\prime }
 + (1-s) \dot{\bf z}_{2 {\rm T}}^{\prime}
 \dot{z}_{10}^{\prime } )^2 \big \}^{1\over 2}
\label{eq:wfase}
\end{eqnarray}
with $ \dot{z}_j^{\mu} ={d z_j^{\mu}(\tau_j) \over d \tau_j}$ as
 in Sec.3.
\par
Let us now introduce
 the momenta  $ p_{j \mu}=-
{\delta S_4^{q \bar{q}}\over \delta \dot{z}_{j}^{\mu}}$:
\begin{eqnarray}
 { p}_{\mu 1}^\prime
 & =& \dot{{ z}}_{\mu 1}^\prime -{4\over 3} \int_0^{\tau_2}
 d\tau_2^\prime D_{\mu \nu}(z_1^\prime-z_2^\prime)
\dot{z}_2^{\prime\nu}- {4\over 3} \int_0^{\tau_1}
 d\tau_2^\prime D_{\mu \nu}(z_1^\prime-z_1^{\prime\prime})
\dot{z}_1^{\prime\prime\nu} +\sigma f_{1\mu}^\prime
\nonumber\\
 {p}_{\mu 2}^\prime & =& \dot{ z}_{\mu 2}^\prime
-{4\over 3} \int_0^{\tau_1}
 d\tau_1^\prime D_{\mu \nu}(z_1^\prime-z_2^\prime)
\dot{z}_1^{\prime\nu}- {4\over 3} \int_0^{\tau_2}
 d\tau_2^\prime D_{\mu \nu}(z_1^\prime-z_1^{\prime\prime})
\dot{z}_2^{\prime\nu}+\sigma f_{2\mu}^\prime
\label{eq:momqq}
\end{eqnarray}
with
\begin{eqnarray}
{\bf u}_s'& = & s \dot{z}_{20}' \dot{\bf z}_{1{\rm T}} +(1-s) \dot{z}_{10}'
\dot{\bf z}_{2 {\rm T}}'\nonumber \\
{\bf f}_1^\prime&=&     \vert {\bf z}_1^\prime-{\bf z}_2^\prime\vert
\int_0^1  ds s  { {\bf u}_{s}^\prime\over [
 \dot{z}_{10}^{2\prime} \dot{z}_{20}^{2\prime} -
 {\bf u}_s^{\prime 2}]^{1\over 2}} \nonumber \\
{\bf f}_2^\prime &=&     \vert {\bf z}_1^\prime-{\bf z}_2^\prime\vert
\int_0^1  ds s (1-s) { {\bf u}_{s}^\prime\over [
 \dot{z}_{10}^{2\prime} \dot{z}_{20}^{2\prime} -
 {\bf u}_s^{\prime 2}]^{1\over 2}}
\nonumber\\
f_{10}^\prime & =& -
  \int_0^{\tau_2} d\tau_2^\prime
 \delta(z_{10}^\prime-z_{20}^\prime) \vert {\bf z}_1^\prime-{\bf
z}_2^\prime \vert
 \int_0^1 ds { \dot{z}_{10}^\prime \dot{z}_{20}^{\prime 2} - (1-s)
\dot{\bf z}_{2{\rm T}}^\prime \cdot {\bf u}_s^\prime\over
[ \dot{z}_{10}^{2\prime} \dot{z}_{20}^{2\prime} -
 {\bf u}_s^{\prime 2}]^{1\over 2}}
\nonumber \\
f_{20}^\prime & =& - \int_0^{\tau_2} d\tau_2'
 \delta(z_{10}'-z_{20}') \vert {\bf z}_1'-{\bf z}_2' \vert
 \int_0^1 ds { \dot{z}_{10}^{'2} \dot{z}_{20}^{'} - s
\dot{\bf z}_{1{\rm T}}' \cdot {\bf u}_s'\over
[ \dot{z}_{10}^{2'} \dot{z}_{20}^{2'} - {\bf u}_s^{'2}]^{1\over 2}}
\label{eq:efdefprim}
\end{eqnarray}
Eq.(\ref{eq:momqq}) cannot be inverted in a closed form, however, we can
invert it by an expansion in $\alpha_s= {g^2 \over 4 \pi }$ and
${\sigma \over m^2}$. At the lowest order we have
\begin{eqnarray}
\dot{z}_1^{\prime \mu}& =&  p_1^{\prime \mu} + {4\over 3} g^2 \int_0^{\tau_2}
 d\tau_2^\prime D_{\mu\nu}(z_1^\prime- z_2^\prime) p_2^{\prime \nu}
+{4\over 3} g^2 \int_0^{\tau_1} d \tau_1^\prime D_{\mu
\nu}(z_1^\prime-z_1^{\prime\prime}) p_1^{\prime \nu}+
 \sigma \tilde{f}_1^{\mu\prime}\nonumber \\
\dot{z}_2^{\prime\mu}& =&  p_2^{\prime\mu} + {4\over 3} g^2 \int_0^{\tau_1}
 d\tau_1^\prime D_{\mu\nu}(z_1^\prime- z_2^\prime) p_1^{\prime \nu}
+{4\over 3} g^2 \int_0^{\tau_2} d \tau_2^\prime D_{\mu
\nu}(z_2^\prime-z_2^{\prime\prime}) p_2^{\prime \nu}+
 \sigma \tilde{f}_2^{\mu\prime}
\label{eq:defv}
\end{eqnarray}
with
\begin{eqnarray}
\tilde{\bf u}_s' & =& s p_{20}' {\bf p}_{1{\rm T}} + (1-s) p_{10}
 {\bf p}_{2{\rm T}}'\nonumber \\
\tilde{\bf f}_1^\prime & =&- \vert
 {\bf z}_1^\prime -{\bf z}_2^\prime \vert \int_0^1 ds\,  s\, {
\tilde{\bf u}_s^\prime \over [p_{10}^{\prime 2}
 p_{20}^{\prime 2} -\tilde{\bf u}_s^{\prime 2}
]^{1\over 2} }\nonumber \\
\tilde{\bf f}_2^\prime & =& -\vert {\bf z}_1^\prime -{\bf z}_2^\prime
 \vert \int_0^1 ds (1-s) {
\tilde{\bf u}_s^\prime \over [p_{10}^{\prime2} p_{20}^{\prime2}
-\tilde{\bf u}_s^{\prime 2}
]^{1\over 2} }\nonumber \\
\tilde{f}_{10}^\prime & =& \vert {\bf z}_1^\prime -{\bf z}_2^\prime
\vert \int_0^1 ds
 {p_{10}^\prime p_{20}^{\prime 2} - (1-s) {\bf p}_{2 {\rm T}}^\prime
\cdot \tilde{\bf
 u}_s^\prime \over [p_{10}^{\prime 2} p_{20}^{\prime 2} -
\tilde{\bf u}_s^{\prime 2}
]^{1\over 2} }\nonumber \\
\tilde{f}_{20}^\prime & =& \vert {\bf z}_1^\prime -{\bf z}_2^\prime
 \vert \int_0^1 ds
 {p_{10}^{\prime 2} p_{20}^{\prime} - s {\bf p}_{1 {\rm T}}^\prime
\cdot \tilde{\bf
 u}_s^\prime \over [p_{10}^{\prime 2} p_{20}^{\prime 2}
-\tilde{\bf u}_s^{\prime 2}
]^{1\over 2} }.
\label{eq:deff}
\end{eqnarray}
Then we can  perform the Legendre trasformation
\begin{eqnarray}
& & \phi^{q \bar{q}}= - \sum_{j=1,2} \int_0^{\tau_j}d\tau_j^{\prime}
 p_j'\cdot \dot{z}_j'
 -S_4^{q\bar{q}}=
\sum_{j=1,2} \int_0^{\tau_j} d\tau_j^\prime  [- {1\over 2}
 (p_j^{\prime 2}
+ m_j^2 )]\nonumber \\
& & + {4\over 3} g^2 \int_0^{\tau_1}
 d\tau_1^\prime \int_0^{\tau_2} d\tau_2^\prime
 D_{\mu\nu}(z_1^\prime- z_2^\prime) p_1^{\prime \mu}
 p_2^{\prime \nu}
+ {2\over 3} g^2 \int_0^{\tau_1}
 d\tau_1^\prime \int_0^{\tau_1} d\tau_1^{\prime\prime}
 D_{\mu\nu}(z_1^\prime- z_1^{\prime\prime}) p_1^{\prime \mu}
 p_1^{\prime \nu}\nonumber \\
& & + {2\over 3} g^2 \int_0^{\tau_2}
 d\tau_2^\prime \int_0^{\tau_2} d\tau_2^{\prime\prime}
 D_{\mu\nu}(z_2^\prime- z_2^{\prime\prime}) p_2^{\prime \mu}
 p_2^{\prime\prime  \nu}\nonumber \\
& &+ \sigma \int_0^{\tau_1} d \tau_1'
\int_0^{\tau_2} d \tau_2' \delta(z_{10}'-z_{20}') \vert {\bf z}_1'
 - {\bf z}_2' \vert \int_0^1  d s ( p_{10}^{'2} p_{20}^{'2}
-\tilde{u}_s^{2\prime})^{1\over 2}
\label{eq:deffi}
\end{eqnarray}
and set
\be
S^{q \bar{q}} = - \sum_{j=1,2} \int_0^{\tau_j} d\tau_j^\prime p_{j\mu} \dot
{z}_j^{\mu} - \phi^{q \bar{q}}.
\ee
In conclusion at the specified order
 the $q \bar{q}$ propagator  can be written as
\begin{eqnarray}
& & G_4^{q \bar{q}}(x_1, x_2; y_1,y_2)  =
 ({i\over 2})^2 \int_0^{\infty} d \tau_1 \int_0^{\infty} d\tau_2
 \int_{y_1}^{x_1} {\cal D}z_1 {\cal D} p_1
\int_{y_2}^{x_2} {\cal D}z_2 {\cal D} p_2 \exp i\big \{
\sum_{j=1,2} \int_0^{\tau_j} d\tau_j' [- p_j' \dot{z}_j'+\nonumber \\
& & + {1\over 2}
 (p_j^{'2} - m_j^2 )]
- {4\over 3} g^2 \int_0^{\tau_1}
 d\tau_1^\prime \int_0^{\tau_2} d\tau_2^\prime
 D_{\mu\nu}(z_1^\prime- z_2^\prime) p_1^{\prime \mu}
 p_2^{\prime \nu}\nonumber \\
& & - {2\over 3} g^2 \int_0^{\tau_1}
 d\tau_1^\prime \int_0^{\tau_1} d\tau_1^{\prime\prime}
 D_{\mu\nu}(z_1^\prime- z_1^{\prime\prime}) p_1^{\prime \mu}
 p_1^{\prime \nu}
- {2\over 3} g^2 \int_0^{\tau_2}
 d\tau_2^\prime \int_0^{\tau_2} d\tau_2^{\prime\prime}
 D_{\mu\nu}(z_2^\prime- z_2^{\prime\prime}) p_2^{\prime \mu}
 p_2^{\prime\prime  \nu}\nonumber \\
& & - \sigma \int_0^{\tau_1} d \tau_1'
\int_0^{\tau_2} d \tau_2' \delta(z_{10}'-z_{20}') \vert {\bf z}_1'
 - {\bf z}_2' \vert \int_0^1  d s ( p_{10}^{'2} p_{20}^{'2}
-\tilde{u}_s^{2\prime})^{1\over 2}+\dots \big \}.
\label{eq:defgult}
\end{eqnarray}
In Eq.(\ref{eq:defgult}) one can proceed as in Eqs.
(\ref{eq:discrprop})--(\ref{eq:cont2}) and one arrives to the   BS
equation (\ref{eq:bspur})
 in the momentum space with a kernel given by
\begin{eqnarray}
& & \tilde{I}(p_1,p_2, p_1^\prime, p_2^\prime) = \int d^4 z_1
 d^4 z_2 \int {d^4 k_1 d^4 k_2\over (2 \pi)^8} \int d^4 z_1^\prime
 d^4 z_2^\prime e^{-i (p_1 -k_1) z_1} e^{-i
(p_2-k_2)z_2}\nonumber \\
& & \Big \{  {16 \over 3} g^2 D_{\mu\nu}({z_1+ z_1^\prime \over 2} -
 {z_2 +z_2^\prime \over 2})k^{\mu}_1 k^{\nu}_2 + \nonumber \\
& & \quad \quad  + 4 \sigma
\delta({z_{10}^\prime +z_{10} \over 2}-{z_{20} + z_{20}^\prime \over
2}) \vert { {\bf z}_1 +{\bf z}_1^\prime \over 2} -{ {\bf z}_2 +
 {\bf z}_2^\prime \over 2}\vert \int_0^1 ds (k_{10} k_{20} - \tilde{u}_s
)^{1\over 2} \Big \} e^{i (p_1^\prime-k_1) z_1^\prime}
e^{i (p_2^\prime-k_2) z_2^\prime}
\label{eq:trault}
\end{eqnarray}
Calculating the integrals in Eq.(\ref{eq:trault}),
   using the definition (\ref{eq:capdef}) and factorizing the
 four--momentum conservation $\delta$
 we arrive to a BS equation of the type (\ref{eq:momrid}) with a
kernel given by (\ref{eq:pertus})-(\ref{eq:kerndef}). \par
 The particular
 Weyl ordering in Eqs.(\ref{eq:trault}), (\ref{eq:bsform}),
(\ref{eq:kerndef}) corresponds to have interpreted (\ref{eq:loop})
 at the discrete level as
\be
W\simeq \langle {\rm Tr P} \prod_{s} \exp i (x_s^{\mu}- x_{s-1}^{\mu})
 A_{\mu} ({x_s + x_{s-1} \over 2}) \rangle
\ee
Notice that $\int_{x_s}^{x_{s+1}} d x^{\mu} A_{\mu} (x) $ differs
 from only $ i (x_s^{\mu} -x_{s-1}^{\mu}) A_{\mu} ({x_s+x_{s-1} \over 2}) $
 by terms of the order $ O[(x_s-x_{s-1})^3] = O(\varepsilon^{3\over
2})$.

\section{Salpeter potential and relativistic flux tube model}

Let us  now consider  the potential to be used in the Salpeter equation
 or in the semirelativistic Schr\"odinger equation and
 corresponding  to the kernel $\hat{I}$ given in
Eqs.(\ref{eq:kern1})--(\ref{eq:kerndef}). \par
 In the scalar model and
 in the center of mass system,
the standard  relation occurring between the BS kernel
 and the potential reads
\begin{equation}
\langle {\bf k} \vert V\vert {\bf k}^\prime\rangle
={1\over (2 \pi )^3}  {1\over 4 \sqrt{ w_1({\bf k})
 w_2({\bf k}) w_1({\bf k}^\prime)  w_2({\bf k}^\prime)}}
 \hat{I}_{\rm inst} ({\bf k}, {\bf k}^\prime),
\label{eq:relpot}
\end{equation}
where $w_j({\bf k})= \sqrt{m_j^2 +{\bf k}^2}$
 and $\hat{I}_{\rm inst}$ denotes the so called instantaneous
 kernel. Precisely $\hat{I}_{\rm inst}$ is obtained
  from $\hat{I}$
 replacing $p_{j0}$ and $p_{j0}^\prime$ by  appropriate
 functions of ${\bf p}_j$ and ${\bf p}_j^\prime$.
 The simplest prescription  corresponds to take
\begin{equation}
p_{j0}=p_{j0}^\prime = { w_j({\bf k})+ w_j({\bf k}^\prime) \over 2}
\label{eq:rep}
\end{equation}
In the Coulomb gauge the resulting potential is
\begin{eqnarray}
& & \langle {\bf k} \vert V\vert {\bf k}^\prime \rangle =
 \rho_1 \rho_2 \Big \{
 - {1\over 2 \pi^2} {4 \over 3} \alpha_s \Big [ {1\over ({\bf
k}^\prime
 -{\bf k})^2} + {1\over  q_{10} q_{20} ({\bf k}^\prime-{\bf k})^2}
\big [ {\bf q}^2 + {\big ( ({\bf k} -{\bf k}^\prime) \cdot {\bf q}\big
)^2
\over ({\bf k}^\prime- {\bf k})^2 }\big ] +\nonumber \\
& &+ {1\over (2 \pi)^3 } \int d^3 {\bf r} e^{i ({\bf k}^\prime -{\bf
k})\cdot {\bf r}}
 \, {\sigma r \over 2}
 {1\over q_{10} +q_{20}} [ {q_{20}\over q_{10}}
 \sqrt{ q_{10}^2 -{\bf q}_{\rm T}^2}
+ {q_{10}\over q_{20}}  \sqrt{q_{20}^2 -{\bf q}^2_{\rm T}} +\nonumber \\
& +& { q_{10} q_{20} \over \vert {\bf q}_{\rm T}\vert }
 ( {\rm arcsin }{ \vert {\bf q}_{\rm T} \vert \over
 \vert q_{10}\vert } + {\rm arcsin } { \vert {\bf q}_{\rm T}\vert
 \over \vert q_{20}\vert } )] + \dots   \Big \}
\label{eq:potfin}
\end{eqnarray}
with ${\bf q} = { {\bf k} +{\bf k}^\prime \over 2} $ and
 $q_{j0} = { w_j({\bf k}) + w_j({\bf k}^\prime) \over 2}$
 and $ \rho_j = { q_{j0} \over \sqrt{ w_j({\bf k}) w_j({\bf k})}} $.
The potential  (\ref{eq:potfin}) corresponds
 to a particular ordering prescription in the  hamiltonian
  (\ref{eq:gamma}) which, as already noticed, is identical
 to  the hamiltonian of the relativistic flux tube model
\cite{flux,olssvar}. Obviously, by an expansion in ${ {\bf q}^2\over
m^2}$  one obtains also the semirelativistic potential derived in
references \cite{BCP94,BP95}. We notice however that,
    due to the substitution (5.2)  and then to the occurrence of the factors
 $\rho_1 \rho_2$ in (\ref{eq:potfin}),  the ordering prescription
  does  not simply coincide
 with the  Weyl prescription given in \cite{BP95}.

\section{Conclusions}
In conclusion we have established  that it is possible to extend the
 Wilson loop method, used to  obtain the semirelativistic potential,
  to the derivation of a Bethe--Salpeter
equation.  We stress that this amounts to show that it is
possible to obtain the BS kernel  from more fundamental arguments
 than those usually used in the phenomenological application
 of the BS equation. The result is strictly related to the
relativistic flux tube model
and to the semirelativistic
potential  for heavy quarks. Some additional remarks are in order.
\begin{itemize}
\begin{enumerate}
\item Due to the difficulty in solving directly the BS--equation
  the general strategy  in the application of such equation
 should be this:    first solve the three--dimensional Salpeter
  equation for the potential (5.3), i.e. the
 eigenvalue equation for the Hamiltonian (1.9);
 then  evaluate
 the ``retardation correction''  by some kind of iterative
method as it is usually done in the positronium case \cite{pos}
 or we have done in \cite{retnos}.
\item In  Equation (3.12) or (3.17), $G_2$ stands for the complete
  one particle propagator which  in principle  should be given
 by (3.5) or its counterpart  for the QCD case. Due  to
 the absence  of a closed path in (3.5), in the present context
 such quantity  should
 be consistently evaluated  simply
by its perturbative  expansion  in contrast
 with what sometimes  supposed.
\item The assumption  of the approximation (1.4) could seem
  unjustified in a relativistic treatment. Notice however
 that the important point in QCD is to have  some zero order
 extimate  of the interesting  quantities  in a formalism
 which already  provide confinement  and then to proceed by
 subsequent corrections.
\item Finally we notice
 that the kernel (\ref{eq:bsform}) is highly singular for
$ {\bf k}^\prime = {\bf k} $, due to the occurrence of the factor $r$
 under the Fourier transform.
This may be  inconvenient  in numerical calculations and  makes
some equation ill defined. The problem is  the same that occurres
 with the linear potential, if one works in the momentum
representation,
 and it
is obviously related to confinement.
 Our philosophy is that one
should introduce an appropriate infrared
regularization (e.g. make the substitution
$ r \rightarrow r e^{- \lambda r}= {\partial^2 \over \partial
\lambda^2} {e^{-\lambda r}\over r} $)
and take the limit $ \lambda \rightarrow
0$
only  at an advanced stage of the calculation.
\end{enumerate}
\end{itemize}

\appendix
\section { }
As an example let us  derive the contribution in
(\ref{eq:kerntant}) corresponding to
   the crossed
 diagram in Fig.1.
Expanding $L_1$  in (3.11)
 we have
\begin{eqnarray}
& & G_4(x_1, x_2; y_1, y_2) = G_2(x_1-y_1) G_2(x_2-y_2)+
  {i g_1 g_2 \over 4}
 ({1 \over 2})^2 \int_0^{\infty} d \tau_1'
 \int_{\tau_1'}^{\infty} d \tau_1 \int_0^{\infty} d \tau_2'
\int_{\tau_2'}^{\infty} d \tau_2 \nonumber\\
& & \int d^4z_1' \int d^4z_2'
\int_{z_1'}^{x_1}  {\cal D} z_1
\int_{z_2'}^{x_2} {\cal D} z_2 \int_{y_1}^{z_1'} {\cal D} z_1
 \int_{y_2}^{z_2'} {\cal D} z_2\,
 D(z_1'-z_2') \nonumber \\
& &  \exp {i\over 2}
 \big \{ \int_{\tau_1'}^{\tau_1} d \tau_1'' [ -\dot{z}_1^{''2}-m_1^2
  +{g_1^2 \over 2} \int_{\tau_1'}^{\tau_1''} d \tau_1'''
D(z_1''-z_1''')] +
 \int_{\tau_2'}^{\tau_2} d \tau_2'' [ - \dot{z}_2^{''2}-m_2^2
 +{g_2^2 \over 2} \int_{\tau_2'}^{\tau_2''} d \tau_2'''
\nonumber \\
& & D(z_2''-z_2''')]\big \}
\times
\exp{i\over 2} \big \{\int_{0}^{\tau_1'} d \tau_1'' [ - \dot{z}_1^{''2}-m_1^2
 +{g_1^2 \over 2} \int_{0}^{\tau_1''} d \tau_1'''
D(z_1''-z_1''')]
+\int_{0}^{\tau_2'} d \tau_2'' [  -\dot{z}_2^{''2}-m_2^2+
\nonumber \\
& &\quad  +{g_2^2 \over 2} \int_{0}^{\tau_2''} d \tau_2'''
D(z_2''-z_2''')]
 +{g_1 g_2 \over 2} \int_0^{\tau_1'} d \tau_1''
 \int_0^{\tau_2'} d \tau_2'' D(z_1''-z_2'')\big \} \times \nonumber \\
& &
\big \{ 1+  {g_1^2 \over 2}
 \int_{\tau_1'}^{\tau_1} d \tau_1'' \int_0^{\tau_1'} d \tau_1'''
 D(z_1''-z_1''') +
{g_2^2 \over 2}
 \int_{\tau_2'}^{\tau_2} d \tau_2'' \int_0^{\tau_2'} d \tau_2'''
 D(z_2''-z_2''') +\nonumber \\
& & \quad \quad \quad \quad + {g_1 g_2 \over 2}
\int_0^{\tau_1'} d \tau_1'' \int_{\tau_2'}^{\tau_2} d \tau_2''
 D(z_1''-z_2'')+\dots \big \}.
\label{eq:app1}
\end{eqnarray}
The contribution  corresponding
to  the crossed diagram (CD) is that coming  from the last
term in  inside the curl bracket in
 (\ref{eq:app1}). We can write
\begin{eqnarray}
& & {\rm CD}= {i g_1 g_2 \over 4} {g_1 g_2 \over 4}
 ({1 \over 2})^2 \int_0^{\infty} d \tau_1'
 \int_{\tau_1'}^{\infty} d \tau_1 \int_0^{\infty} d \tau_2'
\int_{\tau_2'}^{\infty} d \tau_2
\int d^4z_1' \int d^4z_2' \nonumber \\
& & \int_{z_1'}^{x_1}  {\cal D} z_1
\int_{z_2'}^{x_2} {\cal D} z_2 \int_{y_1}^{z_1'} {\cal D} z_1
 \int_{y_2}^{z_2'} {\cal D} z_2\,
\int_0^{\tau_1^\prime} d\tau_1^{\prime\prime}
\int_{\tau_2^\prime}^{\tau_2} d\tau_2^{\prime\prime}
D(z_1^{\prime\prime}-z_2^{\prime\prime})
 D(z_1'-z_2') \nonumber \\
& &  \exp {i\over 2}
 \big \{ \int_{\tau_1'}^{\tau_1} d \tau_1'' [ -\dot{z}_1^{''2}-m_1^2
  +{g_1^2 \over 2} \int_{\tau_1'}^{\tau_1''} d \tau_1'''
D(z_1''-z_1''')] +
 \int_{\tau_2'}^{\tau_2} d \tau_2'' [ - \dot{z}_2^{''2}-m_2^2
 +{g_2^2 \over 2} \int_{\tau_2'}^{\tau_2''} d \tau_2'''
\nonumber \\
& & D(z_2''-z_2''')]\big \}
\times
\exp{i\over 2} \big \{\int_{0}^{\tau_1'} d \tau_1'' [ - \dot{z}_1^{''2}-m_1^2
 +{g_1^2 \over 2} \int_{0}^{\tau_1''} d \tau_1'''
D(z_1''-z_1''')]
+\int_{0}^{\tau_2'} d \tau_2'' [  -\dot{z}_2^{''2}-m_2^2+
\nonumber \\
& &\quad  +{g_2^2 \over 2} \int_{0}^{\tau_2''} d \tau_2'''
D(z_2''-z_2''')]
 +{g_1 g_2 \over 2} \int_0^{\tau_1'} d \tau_1''
 \int_0^{\tau_2'} d \tau_2'' D(z_1''-z_2'')\big \}=\nonumber \\
& & = -{g_1^2 g_2^2 \over 16} ({1 \over 2})^2
\int d^4 z_1^{\prime\prime} \int d^4 z_2^{\prime\prime}
 \int d^4 z_1^{\prime} \int d^4 z_2^\prime D(z_1^\prime-z_2^\prime)
 D(z_1^{\prime\prime}-z_2^{\prime\prime})\nonumber \\
& & \int_0^{\infty}
  d\tau_1^{\prime\prime} \int_{\tau_1^{\prime\prime}}^{\infty}
 d\tau_1^\prime \int_{\tau_1^\prime}^{\infty} d\tau_1
\int_0^{\infty} d\tau_2^{\prime} \int_{\tau_2^\prime}^{\infty}
d\tau_2^{\prime\prime} \int_{\tau_2^{\prime\prime}}^{\infty}
d\tau_2\nonumber \\
&& \int_{z_1^\prime}^{x_1} {\cal D}z_1
\int_{z_1^\prime\prime}^{z_1^\prime} {\cal D}z_1
\int_{y_1}^{z_1^{\prime\prime}} {\cal D} z_1
\int_{z_2^{\prime\prime}}^{x_2} {\cal D} z_2
\int_{z_2^\prime}^{z_2^{\prime\prime}} {\cal D} z_2
\int_{y_2}^{z_2^\prime} {\cal D} z_2\nonumber \\
& &   \exp {i\over 2}
 \big \{ \int_{\tau_1\prime}^{\tau_1} d \tau_1^{\prime\prime}
 [ -\dot{z}_1^{\prime\prime2}-m_1^2
  +{g_1^2 \over 2} \int_{\tau_1^\prime}^{\tau_1\prime\prime} d
\tau_1^{\prime\prime\prime}
D(z_1^{\prime\prime}-z_1^{\prime\prime\prime})] +
\nonumber \\
& & + \int_{\tau_1^\prime\prime}^{\tau_1^\prime } d
\tau_1^{\prime\prime\prime}
 [ -\dot{z}_1^{\prime\prime\prime  2}-m_1^2
  +{g_1^2 \over 2} \int_{\tau_1^\prime}^{\tau_1^{\prime\prime}}
d \tau_1^{\prime \prime \prime\prime }
D(z_1^{\prime\prime\prime}-z_1^{\prime\prime\prime\prime})\big \}
\nonumber \\
&&
  \exp {i\over 2}
 \big \{ \int_{\tau_2\prime}^{\tau_2} d \tau_2^{\prime\prime}
 [ -\dot{z}_2^{\prime\prime 2}-m_2^2
  +{g_2^2 \over 2} \int_{\tau_2^\prime}^{\tau_2\prime\prime} d
\tau_2^{\prime\prime\prime}
D(z_2^{\prime\prime}-z_2^{\prime\prime\prime})] +
\nonumber\\
& &  \int_{\tau_2^\prime\prime}^{\tau_2^\prime } d
\tau_2^{\prime\prime\prime}
 [ -\dot{z}_2^{\prime\prime\prime  2}-m_2^2
  +{g_2^2 \over 2} \int_{\tau_2^\prime}^{\tau_2^{\prime\prime}}
d \tau_2^{\prime \prime \prime\prime }
D(z_2^{\prime\prime\prime}-z_2^{\prime\prime\prime\prime})\big
\}\times
\nonumber\\
& &
\exp {i\over 2} \big \{
\int_{0}^{\tau_1^\prime\prime } d
\tau_1^{\prime\prime\prime}
 [ -\dot{z}_1^{\prime\prime\prime  2}-m_1^2
  +{g_1^2 \over 2} \int_{\tau_1^\prime}^{\tau_1^{\prime\prime}}
d \tau_1^{\prime \prime \prime\prime }
D(z_1^{\prime\prime\prime}-z_1^{\prime\prime\prime\prime}+
\nonumber \\
& &
 \int_{0}^{\tau_2^\prime } d
\tau_2^{\prime\prime\prime}
 [ -\dot{z}_2^{\prime\prime\prime  2}-m_2^2
  +{g_2^2 \over 2} \int_{\tau_2^\prime}^{\tau_2^{\prime\prime}}
d \tau_2^{\prime \prime \prime\prime }
D(z_2^{\prime\prime\prime}-z_2^{\prime\prime\prime\prime}+
\nonumber \\
& & + {g_1 g_2\over 2} \int_0^{\tau_1^{\prime\prime}}
d\tau_1^{\prime\prime\prime} \int_0^{\tau_2^\prime}
d\tau_2^{\prime\prime\prime} D(z_1^{\prime \prime\prime}-
 z_2^{\prime \prime\prime})\big \} \times
\nonumber \\
& &
\exp {i\over 2} \Big \{
 {g_1^2 \over 2} \int_{\tau_1^{\prime\prime}}^{\tau_1^{\prime}}
d\tau_1^{\prime\prime \prime} \int_0^{\tau_1^{\prime\prime}}
d\tau_1^{\prime\prime\prime\prime } D(z_1^{\prime \prime\prime}-
 z_1^{\prime \prime\prime\prime})+
 {g_2^2 \over 2} \int_{\tau_2^{\prime\prime}}^{\tau_2}
d\tau_2^{\prime\prime \prime}
\int_{\tau_2^\prime}^{\tau_2^{\prime\prime}}
d\tau_2^{\prime\prime\prime\prime } D(z_2^{\prime \prime\prime}-
 z_2^{\prime \prime\prime\prime})\nonumber \\
& &
{g_1 g_2 \over 2} \int_{\tau_1^{\prime\prime}}^{\tau_1^\prime}
d\tau_1^{\prime\prime \prime} \int_{0}^{\tau_2^\prime}
d\tau_2^{\prime\prime\prime } D(z_1^{\prime \prime\prime}-
 z_2^{\prime \prime\prime})\Big \}.
\label{eq:orrormax}
\end{eqnarray}
In conclusion,
 replacing again the last exponential $L_2$ in (\ref{eq:orrormax})
 by 1,
we have
\begin{eqnarray}
& & {\rm CD}= i g_1^2 g_2^2 \int d^4 z_1^\prime \int d^4 z_2^\prime
 \int d^4 z_1^{\prime\prime} \int d^4 z_2^{\prime\prime}
D(z_1^\prime - z_2^\prime) D(z_1^{\prime\prime}-z_2^{\prime\prime})
 G_2(x_1-z_1^{\prime}) G_2(z_1^\prime -z_1^{\prime\prime})
\nonumber \\
& & G_2(x_2-z_2^{\prime\prime}) G_2(z_2^{\prime\prime}-z_2^{\prime})
 G_4(z_1^{\prime\prime}, z_2^{\prime}; y_1, y_2).
\label{eq:malapp}
\end{eqnarray}
The method can be immediately extended to higher order terms
or iterated on (A.2) expanding even $L_2$.


\begin{references}
\bibitem{LSG} W. Lucha, F.F. Sch\"oberl and
 D. Gromes, {Phys. Rep.} {\bf 200} {127} (1991)
\bibitem{BCP94} N. Brambilla, P. Consoli and G. M. Prosperi, {\it Phys. Rev.}
 {\bf D 50}  (1994) 5878
\bibitem{BP95} N. Brambilla and G.M. Prosperi, {\it Wilson loops,
 $q \bar{q}$ and $ 3 q$ potentials, Bethe--Salpeter equation},
 in Proceedings of the Conference ``Quark Confinement and the Hadron
 Spectrum'', World Scientific Singapore 1995 and quoted references.
\bibitem{sim} Y. Simonov,  HD-THEP-93-16; Y. Simonov,
 {\it Nucl. Phys. } {\bf B324} (1989) 67; Yu.  Simonov, J. Tjon,
 {\it Ann. Phys.} {\bf 228}  (1993) 1
\bibitem{baker} M. Baker, J. Ball and F. Zachariasen, {Phys. Rev} {bf D47}
 (1993) 3021; {Phys. Lett.} {\bf B283} (1992) 360 and references
therein
\bibitem{lan} F. Langouche, D. Roekaerts, E. Tirapegui, {\it
 Functional integration and semiclassical expansion}, (Reidel,
Dordrecht, 1982)
\bibitem{ind} A.M. Polyakov,  {\it Nucl. Phys.} {\bf B164} (1979) 171;
V.S. Dotsenko and S.N. Vergeles, {\it Nucl. Phys.}
 {\bf B169} (1980) 527; R. Brandt, F. Neri and M. Sato, {\it
 Phys. Rev.} {\bf D24} (1981) 879; see also A. Bassetto
 et al., {\it Nucl. Phys.} {\bf B408} (1993) 62 and references therein
 W. Fischler, {\it Nucl. Phys.} {\bf B129}(1977) 157;
T. Appelquist, M. Dine and I.J. Muzinich, {\it Phys. Rev.} {\bf
 D17} (1978) 2074; A. Billoire, {\it Phys. Lett.} {\bf 92B}
 (1980) 343; F. J. Yndurain and S. Titard, {\it Phys. Rev. }
 {\bf D 49} (1994) 6007
\bibitem{flux} C. Olson, M.G. Olsson and K. Williams, {Phys. Rev.} {\bf D45}
 {4307} (1992); N. Brambilla and G.M. Prosperi, {Phys. Rev.} {\bf D47}
 {2107} (1993); M.G. Olsson and K. Williams, {Phys. Rev.} {\bf D48}
 {417} (1993)
\bibitem{olssvar} C. Olson, M. G. Olsson and D. La Course, {\it Phys.
Rev.} {\bf D49} (1994) 4675; M. G. Olsson, {\it Flux tube and
 Heavy quark Symmetry},
 in Proceedings of the Conference ``Quark Confinement and the Hadron
 Spectrum'', World Scientific Singapore 1995 and quoted references.
\bibitem{peskin} M.A. Peskin, {\it Preprint} {\bf
 SLAC/PUB/3273} (1983)
\bibitem{pos} see e.g. T. Murota, {\it Prog. Theor. Phys.} {\bf 95}
(1988) 46
\bibitem{retnos} N. Brambilla and G.M. Prosperi, {\it Phys. Rev.}
 {\bf D46} (1992) 1096; {\bf D 48} (1993) 2360

\end{references}
\end{document}